\documentclass[10pt,twocolumn]{article}
\usepackage[textwidth=480pt,textheight=661.4pt]{geometry}
\usepackage{amsmath}
\usepackage{amssymb}
\usepackage{relsize}
\usepackage{color}
\usepackage{titlesec}
\usepackage[bottom]{footmisc}
\usepackage[numbers,sort&compress]{natbib}
\usepackage{hyperref}
\usepackage{hypernat}
\usepackage{graphicx}
\usepackage{psfrag}
\usepackage{xspace}
\usepackage{charter}
\usepackage{caption2}

\setcaptionwidth{.9\columnwidth}

\titleformat{\section}[block]
  {\large\bfseries}{\thesection.}{.3em}{}

\DeclareMathOperator{\Tr}{Tr}

\def\x{{\bf x}}

\def\Hk{{\cal H}_{\rm kin}}
\def\Hd{{\cal H}_{\rm diff}}
\def\Hp{{\cal H}_{\rm phys}}

\def\hH{\hat{H}}
\def\hO{\hat{O}}

\newcommand{\wsl}{weak~$\!^*\!$~limit\xspace}

\begin{document}
\onecolumn
\pagestyle{empty}
\begin{flushright}
AEI-2006-004\\
hep-th/0601129\\
January 18th, 2006
\end{flushright}
\vskip 7ex

\begin{center}
\begin{minipage}{.87\textwidth}
{\huge\bf Loop and Spin Foam Quantum Gravity:\\[1ex] 
A Brief Guide for Beginners}\\[6ex]
{\large\bf Hermann Nicolai and Kasper Peeters}\\[5ex]
Max-Planck-Institut f\"ur Gravitationsphysik\\
Albert-Einstein-Institut\\
Am M\"uhlenberg 1\\
14476 Golm, GERMANY\\[3ex]
{\tt hermann.nicolai}, 
{\tt kasper.peeters@aei.mpg.de}
\vskip 9ex

{\bf Abstract:}\\[1ex] We review aspects of loop quantum gravity and
spin foam models at an introductory level, with special attention to 
questions frequently asked by non-specialists.
\end{minipage}
\vfill
\begin{minipage}{0.87\textwidth}
\noindent{\smaller\smaller\smaller Contributed article to ``\emph{An assessment of current 
paradigms in the physics of fundamental interactions}'',
  ed.~I. Stamatescu, Springer Verlag.}
\end{minipage}
\end{center}
\twocolumn
\pagestyle{plain}
\section{Quantum Einstein gravity}

The assumption that Einstein's classical theory of gravity can be
quantised non-perturbatively is at the root of a wide variety of
approaches to quantum gravity. The assumption constitutes the basis of
several discrete methods~\cite{Loll:1998aj}, such as dynamical
triangulations and Regge calculus, but it also implicitly underlies
the older Euclidean path integral approach~\cite{Gibbons:1976ue,hawk3}
and the somewhat more indirect arguments which suggest that there may
exist a non-trivial fixed point of the renormalisation
group~\mbox{\cite{wein4,wein3,Lauscher:2001rz}}. Finally, it is the
key assumption which underlies loop and spin foam quantum
gravity. Although the assumption is certainly far-reaching, there is
to date no proof that Einstein gravity cannot be quantised
non-perturbatively, either along the lines of one of the programs
listed above or perhaps in an entirely different way.

In contrast to string theory, which posits that the Einstein-Hilbert
action is only an effective low energy approximation to some other,
more fundamental, underlying theory, loop and spin foam gravity take
Einstein's theory in four spacetime dimensions as the basic starting
point, either with the conventional or with a (constrained) `BF-type' 
formulation.\footnote{In the remainder, we will often follow established 
(though perhaps misleading) custom and summarily refer to this
framework of ideas simply as ``Loop Quantum Gravity'', or LQG for
short.} These approaches are background independent in the sense that
they do not presuppose the existence of a given background metric. 
In comparison to the older geometrodynamics
approach (which is also formally background independent) they make use
of many new conceptual and technical ingredients. A key role is played
by the reformulation of gravity in terms of connections and
holonomies. A related feature is the use of spin networks in three
(for canonical formulations) and four (for spin foams)
dimensions. These, in turn, require other mathematical ingredients,
such as non-separable (`polymer') Hilbert spaces and representations
of operators which are not weakly continuous. Undoubtedly, novel 
concepts and ingredients such as these will be necessary in order 
to circumvent the problems of perturbatively quantised gravity (that 
novel ingredients are necessary is, in any case, not just the point 
of view of LQG but also of most other approaches to quantum gravity).  
Nevertheless, it is important not to lose track of the physical questions 
that one is trying to answer. 

Evidently, in view of our continuing ignorance about the `true theory' of 
quantum gravity, the best strategy is surely to explore all possible avenues. 
LQG, just like the older geometrodynamics approach~\cite{Kiefer:2004gr},
addresses several aspects of the problem that are currently outside
the main focus of string theory, in particular the question of
background independence and the quantisation of geometry. Whereas
there is a rather direct link between (perturbative) string theory and
classical space-time concepts, and string theory can therefore rely on
familiar notions and concepts, such as the notion of a particle and
the S-matrix, the task is harder for LQG, as it must face up right away 
to the question of what an observable quantity is in the absence of a 
proper semi-classical space-time with fixed asymptotics. 

The present text, which is based in part on the companion
review~\cite{kas_lqg}, is intended as a brief introductory and
critical survey of loop and spin foam quantum gravity~\footnote{Whereas
\cite{kas_lqg} is focused on the `orthodox' approach to loop
quantum gravity, to wit the Hamiltonian framework.}, with special
attention to some of the questions that are frequently asked by
non-experts, but not always adequately emphasised (for our taste, at
least) in the pertinent literature. For the canonical formulation of
LQG, these concern in particular the definition and implementation of
the Hamiltonian (scalar) constraint and its lack of uniqueness. An
important question (which we will not even touch on here) concerns the
consistent incorporation of matter couplings, and especially the
question as to whether the consistent quantisation of gravity imposes
any kind of restrictions on them. Establishing the existence of a
semi-classical limit, in which classical space-time and the Einstein
field equations are supposed to emerge, is widely regarded as the main
open problem of the LQG approach. This is also a prerequisite for
understanding the ultimate fate of the non-renormalisable UV
divergences that arise in the conventional perturbative
treatment. Finally, in any canonical approach there is the question
whether one has succeeded in achieving (a quantum version of) full
space-time covariance, rather than merely covariance under
diffeomorphisms of the three-dimensional slices. In~\cite{kas_lqg} we
have argued (against a widely held view in the LQG community) that for
this, it is not enough to check the closure of two Hamiltonian
constraints on diffeomorphism invariant states, but that it is rather
the \emph{off-shell closure} of the constraint algebra that should be
made the crucial requirement in establishing quantum space-time
covariance.

Many of these questions have counterparts in the spin foam approach, 
which can be viewed as a `space-time covariant version' of LQG, and
at the same time as a modern variant of earlier attempts to define
a discretised path integral in quantum gravity. For instance, the 
existence of a semi-classical limit is related to the question whether 
the Einstein-Hilbert action can be shown to emerge in the infrared (long 
distance) limit, as is the case in (2+1) gravity in the 
Ponzano-Regge formulation, cf.~eq.~(\ref{e:6jlarge}). Regarding the 
non-renormalisable UV divergences of perturbative quantum gravity, 
many spin foam practitioners seem to hold the view that there is no 
need to worry about short distance singularities and the like because 
the divergences are simply `not there' in spin foam models, due to 
the existence of an intrinsic cutoff at the Planck scale. However, 
the same statement applies to any regulated quantum field theory (such
as lattice gauge theory) before the regulator is removed, and on 
the basis of this more traditional understanding, one would therefore 
expect the `correct' theory to require some kind of refinement (continuum) 
limit \footnote{Unless quantum gravity is ultimately a \emph{topological}
theory, in which case the sequence of refinements becomes stationary. 
Such speculations have also been entertained in the context of string 
and M~theory.}, or a sum `over all spin foams' (corresponding to the 
`sum over all metrics' in a formal path integral). If one accepts this 
point of view, a key question is whether it is possible to obtain results 
which do not depend on the specific way in which the discretisation and 
the continuum limit are performed (this is also a main question in other 
discrete approaches which work with reparametrisation invariant quantities, 
such as in Regge calculus). On the other hand, the very need to take such 
a limit is often called into question by LQG proponents, who claim that 
the discrete (regulated) model correctly describes physics at the Planck
scale. However, it is then difficult to see (and, for gravity in (3+1)
dimensions has not been demonstrated all the way in a single example) 
how a classical theory with all the requisite properties, and in particular 
full space-time covariance, can emerge at large distances. Furthermore, 
without considering such limits, and in the absence of some other unifying 
principle, one may well remain stuck with a multitude of possible models, 
whose lack of uniqueness simply mirrors the lack of uniqueness that comes 
with the need to fix infinitely many coupling parameters in the 
conventional perturbative approach to quantum gravity.

Obviously, a brief introductory text such as this cannot do justice to
the numerous recent developments in a very active field of current
research.  For this reason, we would like to conclude this
introduction by referring readers to several `inside' reviews for
recent advances and alternative points of view, namely
\mbox{\cite{Gambini:1996ik,Thiemann:2001yy,Ashtekar:2004eh}} for the
canonical formulation,
\mbox{\cite{Perez:2004hj,Baez:1999sr,Perez:2003vx}} for spin foams,
and \cite{b_rove1} for both. A very similar point of view to ours has
been put forward
in~\cite{Perez:2005fn,Perez:2006gj}\footnote{However, \cite{Perez:2005fn,Perez:2006gj}
only addresses the so-called \mbox{$m$-ambiguity}, whereas we will
argue that there are infinitely many other parameters which a
microscopic theory of quantum gravity must fix.}. Readers are also
invited to have a look at~\cite{loops05} for an update on the very
latest developments in the subject.

\section{The kinematical Hilbert space of LQG}

\begin{figure}[t]
\psfrag{e1}{$e_1$}
\psfrag{e2}{$e_2$}
\psfrag{eeps}{}
\begin{center}
\vspace{1ex}
\includegraphics[width=.6\columnwidth]{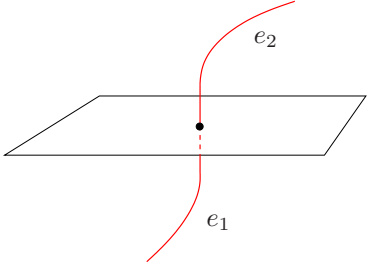}
\caption{LQG employs holonomies and fluxes as elementary conjugate
  variables. When the edge of the holonomy and the two-surface element of
  the flux intersect, the canonical Poisson bracket of the associated
  operators is non-vanishing, and inserts a $\tau$-matrix at the point
  of intersection, cf.~\eqref{e:basic_bracket}.\label{f:canonical_bracket}}
\end{center}
\end{figure}

There is a general expectation (not only in the LQG community) that at
the very shortest distances, the smooth geometry of Einstein's theory
will be replaced by some quantum space or spacetime, and hence the continuum 
will be replaced by some `discretuum'. Canonical LQG does not do away
with conventional spacetime concepts entirely, in that it still relies 
on a spatial continuum~$\Sigma$ as its `substrate', on which holonomies
and spin networks live (or `float') --- of course, with the idea of 
eventually `forgetting about it' by considering abstract spin
networks and only the combinatorial relations between them. On this 
substrate, it takes as the classical phase space variables the holonomies 
of the Ashtekar connection,
\begin{multline}
\label{e:holonomy_def}
h_e [A] = {\cal P} \exp \int_e A_m^a\tau_a {\rm
  d}x^m\,,\\[1ex]\quad\text{with}\quad
A_{m}^a := -\tfrac{1}{2}\epsilon^{abc} \omega_{m bc} +
\gamma\, K_{m}^a\,.
\end{multline}
Here, $\tau_a$ are the standard generators of~SU(2) (Pauli matrices),
but one can also replace the basic representation by a representation of
arbitrary spin, denoted by~$\rho_j (h_e[A])$.
The Ashtekar connection~$A$ is thus a particular linear combination of the
spin connection~$\omega_{m b c}$ and the extrinsic curvature~$K_m^a$
which appear in a standard (3+1) decomposition. The parameter~$\gamma$
is the so-called Barbero-Immirzi parameter. The variable conjugate to
the Ashtekar connection turns out to be the inverse densitised
dreibein~\mbox{$\tilde{E}_a{}^m := e\, e_a{}^m$}. Using this conjugate
variable, one can find the objects which are conjugate to the
holonomies. These are given by integrals of the associated two-form
over two-dimensional surfaces~$S$ embedded in~$\Sigma$,
\begin{equation}
\label{e:flux_def1}
F_S[\tilde E, f] := \int_S \, \epsilon_{mnp} \tilde{E}_a^m\,f^a\,{\rm d}x^n
\wedge {\rm d}x^p\,,
\end{equation}
where $f^a(x)$ is a test function.  This flux vector is indeed
conjugate to the holonomy in the sense described in
figure~\ref{f:canonical_bracket}: if the edge associated to the
holonomy intersects the surface associated to the flux, the Poisson
bracket between the two is non-zero,
\begin{multline}
\label{e:basic_bracket}
\Big\{ (h_e[A])_{\alpha\beta}, \, F_S[\tilde{E},f] \Big\} \\[1ex]
= \pm\, \gamma\, f_a(P) \big(h_{e_1}[A]\,\tau^a\, h_{e_2}[A]\big)_{\alpha\beta} \,,
\end{multline}
where $e = e_1 \cup e_2$ and the sign depends on the relative
orientation of the edge and the two-surface.  This Poisson structure
is the one which gets promoted to a commutator algebra in the quantum
theory.

\begin{figure}[t]
\begin{center}
\psfrag{e1}{\small $e_1$} \psfrag{e2}{\small $\!e_2$}
\psfrag{e3}{\small $\!\!e_3$} \psfrag{e4}{\small $e_4$} \psfrag{e5}{\small
$e_5$} \psfrag{e6}{\small $\!\!e_6$} \psfrag{e7}{\small $e_7$}
\psfrag{e8}{\small $e_8$} \psfrag{n1}{\small $\!\!v_1$} \psfrag{n2}{\small
$v_2$} \psfrag{n3}{\small $v_3$} \psfrag{n4}{\small $v_4$}
\psfrag{n5}{\small $v_5$}
\psfrag{sigma}{\larger\larger$\Sigma$}
\vspace{3ex}
\includegraphics[width=0.8\columnwidth]{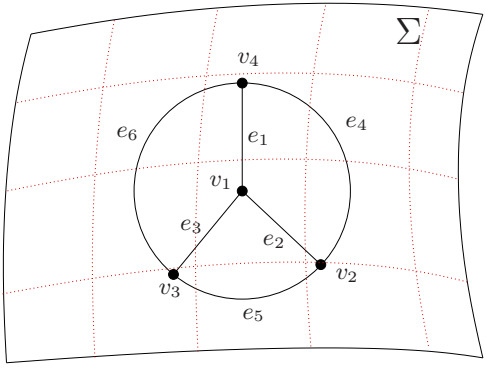}
\end{center}
\caption{A simple spin network, embedded in the spatial
  hypersurface~$\Sigma$. The hypersurface is only present in order to
  provide coordinates which label the positions of the vertices and
  edges. Spin network wave functions only probe the geometry along the
  one-dimensional edges and are insensitive to the geometry
  elsewhere on~$\Sigma$.\label{f:mercedes}}
\end{figure}

Instead of building a Hilbert space as the space of functions over
configurations of the Ashtekar connection, i.e.~instead of
constructing wave-functionals $\Psi[A_m(\x)]$, LQG uses a Hilbert
space of wave functionals which ``probe'' the geometry only on
one-dimensional submanifolds, so-called \emph{spin networks}. The latter
are (not necessarily connected) graphs $\Gamma$ embedded in $\Sigma$
consisting of finitely many edges (links). The wave functionals are
functionals over the space of ho\-lo\-nomies. In order to make them
$\mathbb{C}$-valued, the SU(2) indices of the holonomies have to be
contracted using invariant tensors (i.e.~Clebsch-Gordan
coefficients). The wave function associated to the spin network in
figure~\ref{f:mercedes} is, for instance, given by
\begin{multline}\label{e:threevalent}
\Psi [{\text{fig.\ref{f:mercedes}}}] = 
\Big( \rho_{j_1}(h_{e_1}[A])\Big)_{\alpha_1 \beta_1}\; 
\Big( \rho_{j_2}(h_{e_2}[A])\Big)_{\alpha_2 \beta_2}\;\\[1ex]
\times\Big(\rho_{j_3}(h_{e_3}[A])\Big)_{\alpha_3 \beta_3}\, 
C^{j_1 j_2 j_3}_{\beta_1 \beta_2 \beta_3}  \ldots \,,
\end{multline}
where dots represent the remainder of the graph. The spin labels $j_1,\dots$
must obey the standard rules for the vector addition of angular momenta, but 
otherwise can be chosen arbitrarily. The spin network wave functions $\Psi$ 
are thus labelled by $\Gamma$ (the spin network graph), by the spins $\{j\}$ 
attached to the edges, and the intertwiners $\{C\}$ associated to the 
vertices.

At this point, we have merely defined a space of wave functions in
terms of rather unusual variables, and it now remains to define a
proper Hilbert space structure on them. The discrete kinematical
structure which LQG imposes does, accordingly, \emph{not} come from
the description in terms of holonomies and fluxes. After all, this
very language can also be used to describe ordinary Yang-Mills theory.
The discrete structure which LQG imposes is also entirely different
from the discreteness of a lattice or naive discretisation of space
({\it i.e.\/}~of a finite or countable set). Namely, it arises by
`polymerising' the continuum via an unusual \emph{scalar product}. For
any two spin network states, one defines this scalar product to be
\begin{multline}
\label{e:scalar_product}
\big\langle \Psi_{\Gamma,\{j\},\{C\}} \,\big|\, 
\Psi'_{\Gamma',\{j'\},\{C'\}} \big\rangle \\[1ex]
=\!\begin{cases}
\;\;0 &\text{\!if $\Gamma \neq \Gamma'$}\,, \\[1ex]
\displaystyle\int\!\! \prod_{e_i\in\Gamma} {\rm d}h_{e_i}\;
\, \bar\psi_{\Gamma,\{j\}, \{C\}}\,
\psi'_{\Gamma',\{j'\},\{C'\}} & 
\text{\!if $\Gamma = \Gamma'$} \,,
\end{cases}
\end{multline}
where the integrals~$\int\!{\rm d}h_e$ are to be performed with the
SU(2) Haar measure. The spin network wave functions~$\psi$ depend on
the Ashtekar connection only through the holonomies. The \emph{kinematical
Hilbert space} $\Hk$ is then defined as the completion of the space of spin 
network wave functions w.r.t.~this scalar product~\eqref{e:scalar_product}. 
The topology induced by the latter is similar to the discrete topology
(`pulverisation') of the real line with countable unions of points as
the open sets. Because the only notion of `closeness' between two
points in this topology is whether or not they are coincident, whence
\emph{any} function is continuous in this topology, this raises the
question as to how one can recover conventional notions of continuity
in this scheme.

The very special choice of the scalar product~\eqref{e:scalar_product}
leads to representations of operators which need not be weakly continuous:
this means that expectation values of operators depending on some
parameter do not vary continuously as these parameters are varied.
Consequently, the Hilbert space does not admit a countable basis,
hence is \emph{non-separable}, because the set of all spin network
graphs in~$\Sigma$ is uncountable, and non-coincident spin networks
are orthogonal w.r.t.~\eqref{e:scalar_product}.  Therefore, any
operation (such as a diffeomorphism) which moves around graphs
continuously corresponds to an uncountable sequence of mutually
orthogonal states in~$\Hk$. That is, no matter how `small' the
deformation of the graph in~$\Sigma$, the associated elements of~$\Hk$
always remain a finite distance apart, and consequently, the
continuous motion in `real space' gets mapped to a highly
discontinuous one in~$\Hk$.  Although unusual, and perhaps
counter-intuitive, as they are, these properties constitute a
cornerstone for the hopes that LQG can overcome the seemingly
unsurmountable problems of conventional geometrodynamics: if the
representations used in LQG were equivalent to the ones of
geometrodynamics, there would be no reason to expect LQG not to end up
in the same quandary.

Because the space of quantum states used in LQG is very different from
the one used in Fock space quantisation, it becomes non-trivial to see
how semi-classical `coherent' states can be constructed, and how a
smooth classical spacetime might emerge. In simple toy examples, such
as the harmonic oscillator, it has been shown that the LQG Hilbert
space indeed admits states (complicated linear superpositions) whose
properties are close to those of the usual Fock space coherent
states~\cite{Ashtekar:2002sn}. In full \mbox{(3+1)-dimensional} LQG,
the classical limit is, however, far from understood (so far only
kinematical coherent states are
known~\cite{Thiemann:2000bw,Thiemann:2000ca,Thiemann:2000bx,Thiemann:2000by,Thiemann:2002vj,Sahlmann:2001nv},
i.e.~states which do not satisfy the quantum constraints). In
particular, it is not known how to describe or approximate classical
spacetimes in this framework that `look' like, say, Minkowski space,
or how to properly derive the classical Einstein equations and their
quantum corrections. A proper understanding of the semi-classical
limit is also indispensable to clarify the connection (or lack
thereof) between conventional perturbation theory in terms of Feynman
diagrams, and the non-perturbative quantisation proposed by LQG.

However, the truly relevant question here concerns the structure (and
definition!) of \emph{physical} space and time. This, and not the
kinematical `discretuum' on which holonomies and spin networks
`float', is the arena where one should try to recover familiar and
well-established concepts like the Wilsonian renormalisation group,
with its \emph{continuous} `flows'. Because the measurement of lengths
and distances ultimately requires an operational definition in terms
of appropriate matter fields and states obeying the physical state
constraints, `dynamical' discreteness is expected to manifest itself
in the spectra of the relevant physical observables. Therefore, let us
now turn to a discussion of the spectra of three important operators
and to the discussion of physical states.

\section{Area, volume and the Hamiltonian}

In the current setup of LQG, an important role is played by two 
relatively simple operators: the `area operator' measuring the area 
of a two-dimensional surface~$S\subset\Sigma$, and the `volume operator' 
measuring the volume of a three-dimensional subset~$V\subset\Sigma$.
The latter enters the definition of the Hamiltonian constraint in
an essential way. Nevertheless, it must be emphasised that the
area and volume operators are \emph{not} observables in the Dirac
sense, as they do not commute with the Hamiltonian. To construct
\emph{physical} operators corresponding to area and volume is more
difficult and would require the inclusion of matter (in the form of
`measuring rod fields').

\begin{figure}[t]
\begin{center}
\psfrag{c}{}
\vspace{3ex}
\includegraphics[width=.6\columnwidth]{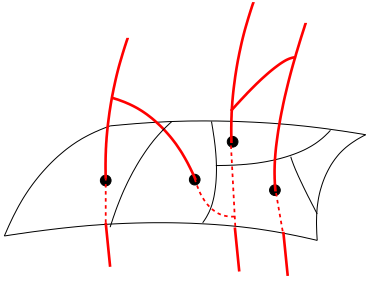}
\end{center}
\caption{The computation of the spectrum of the area operator involves
  the division of the surface into cells, such that at most one edge
  of the spin network intersects each given cell.\label{f:area_subdiv}}
\end{figure}

\begin{figure*}[t]
\setcaptionwidth{.9\textwidth}
\psfrag{1}{\smaller\smaller 1}
\psfrag{2}{\smaller\smaller 2}
\psfrag{3}{\smaller\smaller 3}
\psfrag{4}{\smaller\smaller 4}
\begin{center}
\vspace{3ex}
\includegraphics[width=.8\textwidth]{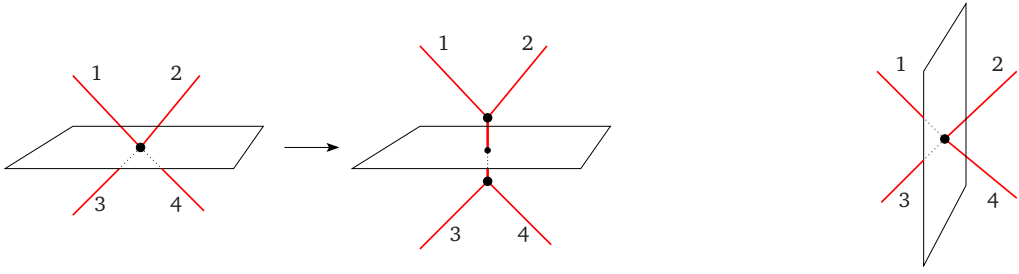}
\end{center}
\caption{The action of the area operator on a node with intertwiner
$C^{j_1 j_2 k}_{\alpha_1 \alpha_2 \beta} C^{j_3 j_4 k}_{\alpha_3
\alpha_4 \beta}$. Whether or not this action is diagonal depends on
the orientation of the surface associated to the area operator.  In
the figure on the left, the location of the edges with respect to the
surface is such that the invariance of the Clebsch-Gordan coefficients
can be used to evaluate the action of the area operator. The result
can be written in terms of a ``virtual'' edge. In the figure on the
right, however, this is not the case, a recoupling relation is needed,
and the spin network state is not an eigenstate of the corresponding
area operator.\label{f:area_nondiag}}
\end{figure*}

The area operator is most easily expressed as
\begin{multline}
A_{S}[g] = \int_{S} \, \sqrt{{\rm d}F^a \cdot {\rm d}F^a}\,,\\[1ex]
\text{with}\quad
{\rm d}F_a := \epsilon_{mnp} \tilde{E}_a^m {\rm d}x^n \wedge {\rm d}x^p
\end{multline}
(the area element is here expressed in terms of the new `flux variables'
$\tilde{E}_a^m$, but is equal to the standard expression
${\rm d}F_a := \epsilon_{abc} e_m{}^b e_n{}^c {\rm d}x^m \wedge {\rm d}x^n$).
The next step is to re-write this area element in terms of the spin
network variables, in particular the momentum~$\tilde{E}_a{}^m$
conjugate to the Ashtekar connection.  In order to do so, we subdivide
the surface into infinitesimally small surfaces~$S_I$ as in
figure~\ref{f:area_subdiv}.  Next, one approximates the area by a
Riemann sum (which, of course, converges for well-behaved
surfaces~$S$), using
\begin{equation}
\int_{S_I} \, \sqrt{{\rm d}F^a \cdot {\rm d}F^a} \approx
\sqrt{{F^a_{S_I}[\tilde{E}]\, F^a_{S_I}[\tilde{E}] }}\,.
\end{equation}
This turns the operator into the expression
\begin{equation}
\label{Area}
A_{S}[\tilde{E}_m^a] 
= \lim_{N \rightarrow \infty} \sum_{I=1}^N
\sqrt{{F^a_{S_I}[\tilde{E}]\, F^a_{S_I}[\tilde{E}] }} \, . 
\end{equation}
If one applies the operator~(\ref{Area}) to a wave function associated
with a fixed graph $\Gamma$ and refines it in such a way that each
elementary surface~$S_I$ is pierced by only \emph{one} edge of the
network, one obtains, making use of~\eqref{e:basic_bracket} twice,
\begin{equation} 
\label{Area-net}
\hat{A}_S  \Psi  = 8 \pi l_p^2 \gamma \sum_{p=1}^{\#\text{edges}}
\sqrt{j_p(j_p + 1)}\, \Psi  \, .
\end{equation}
These spin network states are thus eigenstates of the area
operator. The situation becomes considerably more complicated for wave
functions which contain a spin network vertex which lies in the
surface~$S$; in this case the area operator does not necessarily act
diagonally anymore (see
figure~\ref{f:area_nondiag}). Expression~\eqref{Area-net} lies at the
core of the statement that areas are quantised in LQG.

The construction of the volume operator follows similar logic,
although it is substantially more involved. One starts with the
classical expression for the volume of a three-dimensional region
$\Omega\subset\Sigma$,
\begin{equation} 
\label{e:vol}
V(\Omega) = \int_\Omega\! {\rm d}^3x \sqrt{\left|\frac{1}{3!} \epsilon_{abc} 
\epsilon^{mnp}\tilde{E}_m^a \tilde{E}_n^b \tilde{E}_p^c\right|}\,.
\end{equation}
Just as with the area operator, one partitions~$\Omega$ into small
cells $\Omega = \cup_I \Omega_I$, so that the integral can be replaced
with a Riemann sum. In order to express the volume element in terms of
the canonical quantities introduced before, one then again approximates
the area elements ${\rm d}F^a$ by the small but finite area operators
$F^a_S[\tilde{E}]$, such that the volume is obtained as the limit of a
Riemann sum
\begin{equation}
\label{e:vol-sum0}
V(\Omega) = \lim_{N \rightarrow \infty} \sum_{I=1}^N 
\sqrt{  \left| \frac1{3!} \epsilon_{abc}
F^a_{S^1_I}[\tilde{E}]\, F^b_{S^2_I}[\tilde{E}]\,
F^c_{S^3_I}[\tilde{E}]  \right|   } \,.
\end{equation}
The main problem is now to choose appropriate surfaces~$S_{1,2,3}$ in
each cell. This should be done in such a way that the r.h.s. of
\eqref{e:vol-sum0} reproduces the correct classical value. For
instance, one can choose a point inside each cube $\Omega_I$, then
connect these points by lines and `fill in' the faces. In each cell
$\Omega_I$ one then has three lines labelled by $a=1,2,3$; the surface
$S^a_I$ is then the one that is traversed by the $a$-th line. With
this choice it can be shown that the result is insensitive to small
`wigglings' of the surfaces, hence independent of the shape of
$S^a_I$, and the above expression converges to the desired result.
See~\cite{Brunnemann:2005in,Meissner:2005mx} for some recent results
on the spectrum of the volume operator.

The key problem in canonical gravity is the definition and implementation
of the Hamiltonian (scalar) constraint operator, and the verification
that this operator possesses all the requisite properties. The latter
include (quantum) space-time covariance as well as the existence of a 
proper semi-classical limit, in which the classical Einstein equations
are supposed to emerge. It is this operator which replaces the Hamiltonian 
evolution operator of ordinary quantum mechanics, and encodes all the 
important dynamical information of the theory (whereas the Gauss and 
diffeomorphism constraints are merely `kinematical'). More specifically, 
together with the kinematical constraints, it defines the \emph{physical 
states} of the theory, and thereby the physical Hilbert space~$\Hp$
(which may be separable~\cite{Fairbairn:2004qe}, even is~$\Hk$ is not).

To motivate the form of the \emph{quantum Hamiltonian} one starts with
the classical expression, written in loop variables. To this aim one
re-writes the Hamiltonian in terms of Ashtekar variables, with the result 
\begin{multline}
\label{e:clH}
H[N] = \int_\Sigma\!{\rm d}^3x\, 
N \frac{\tilde{E}_a^m \tilde{E}_b^n}{\sqrt{\det \tilde{E}}} 
\Big( \epsilon^{abc} F_{mnc} \\[1ex] - \frac{1}{2} (1+\gamma^2) K_{[m}{}^a
  K_{n]}{}^b \Big)\,.
\end{multline}
For the special values~$\gamma = \pm i$, the last term drops out, and
the Hamiltonian simplifies considerably. This was indeed the value
originally proposed by Ashtekar, and it would also appear to be the
natural one required by local Lorentz invariance (as the Ashtekar
variable is, in this case, just the pullback of the four-dimensional
spin connection).  However, imaginary~$\gamma$ obviously implies that
the phase space of general relativity in terms of these variables
would have to be complexified, such that the original phase space
could be recovered only after imposing a reality constraint. In order
to avoid the difficulties related to quantising this reality
constraint, $\gamma$ is now usually taken to be real.  With this
choice, it becomes much more involved to rewrite~\eqref{e:clH} in
terms of loop and flux variables. 

\section{Implementation of the constraints}
\label{s:constraints}

In canonical gravity, the simplest constraint is the Gauss
constraint. In the setting of LQG, it simply requires that the SU(2)
representation indices entering a given vertex of a spin network enter
in an SU(2) invariant manner. More complicated are the diffeomorphism
and Hamiltonian constraint. In LQG these are implemented in two
entirely different ways. Moreover, the implementation of the
Hamiltonian constraint is not completely independent, as its very
definition relies on the existence of a subspace of diffeomorphism 
invariant states.

Let us start with the diffeomorphism constraint. Unlike in
geometrodynamics, one cannot immediately write down formal states
which are manifestly diffeomorphism invariant, because the spin
network functions are not supported on all of~$\Sigma$, but only on
one-dimensional links, which `move around' under the action of a
diffeomorphism. A formally diffeomorphism invariant state is obtained 
by `averaging' over the diffeomorphism group, and more specifically
by considering the formal sum
\begin{equation}
\label{e:diffintegral}
\eta(\Psi)[A] :=
\sum_{\phi\in {\text{Diff}(\Sigma|\Gamma)}}
\Psi_{\Gamma}[A\circ\phi]\, .
\end{equation}
Here Diff($\Sigma|\Gamma$) is obtained by dividing out the
diffeomorphisms leaving invariant the graph~$\Gamma$. Although this is
a continuous sum which might seem to be ill-defined, it can be given a
mathematically precise meaning because the unusual scalar
product~\eqref{e:scalar_product} ensures that the inner product
between a state and a diffeomorphism-averaged state,
\begin{equation}
\langle \eta(\Psi_{\Gamma'})\, |\, \Psi_{\Gamma}\rangle =
\sum_{\phi\in {\text{Diff}(\Sigma|\Gamma')}}
\langle \phi^* \circ \Psi_{\Gamma'}\,|\, \Psi_{\Gamma} \rangle
\end{equation}
consists at most of a \emph{finite} number of terms. It is this fact
which ensures that~$\langle \eta(\Psi_{\Gamma})|$ is indeed well-defined 
as an element of the space dual to the space of spin networks (which
is dense in $\Hk$).
In other words, although $\eta(\Psi)$ is certainly outside of $\Hk$, it 
does make sense \emph{as a distribution}. On the space of diffeomorphism
averaged spin network states (regarded as a subspace of a distribution
space) one can now again introduce a Hilbert space structure `by dividing 
out' spatial diffeomorphisms, namely
\begin{equation}
\langle\!\langle \eta(\Psi)| \eta (\Psi')\rangle\!\rangle :=
\langle \eta(\Psi)| \Psi' \rangle\,.
\end{equation}
The completion by means of this scalar product defines the space $\Hd$;
but note that $\Hd$ is \emph{not} a subspace of $\Hk$!

As we said above, however, it is the Hamiltonian constraint which 
plays the key role in canonical gravity, as it this operator which
encodes the dynamics. Implementing this constraint on $\Hd$ or some 
other space is fraught with numerous choices and ambiguities,
inherent in the construction of the quantum Hamiltonian as well as the
extraordinary complexity of the resulting expression for the
constraint operator~\cite{Borissov:1997ji}. The number of ambiguities
can be reduced by invoking independence of the spatial
background~\cite{Thiemann:2001yy}, and indeed, without making such
choices, one would not even obtain sensible expressions.
In other words, the formalism is partly `on-shell' in that 
the very existence of the (unregulated) Hamiltonian
constraint operator depends very delicately on its `diffeomorphism
covariance', and the choice of a proper `habitat', on which it is
supposed to act in a well defined manner. A further source of
ambiguities, which, for all we know, has not been considered in the
literature so far, consists in possible $\hbar$-dependent `higher
order' modifications of the Hamiltonian, which might still be compatible 
with all consistency requirements of LQG.

In order to write the constraint in terms of only holonomies and
fluxes, one has to eliminate the inverse square root $\tilde{E}^{-1/2}$
in (\ref{e:clH}) as well as the
extrinsic curvature factors. This can be done through a number of
tricks found by Thiemann~\cite{Thiemann:1996aw}. The vielbein
determinant is eliminated using
\begin{equation}
\label{e:trick1}
  \epsilon_{mnp} \epsilon^{abc} \tilde{E}^{-1/2} 
  \tilde{E}_b{}^n \tilde{E}_c{}^p =
  \frac1{4\gamma} \Big\{ A_m{}^a (\x) , V \Big\}\,.
\end{equation}
where $V\equiv V(\Sigma)$ is the total volume, cf.~(\ref{e:vol}).
The extrinsic curvature is eliminated by writing it as
\begin{multline}
\label{e:trick2}
K_m{}^a (\x )= \frac{1}{\gamma}\big\{ A_m{}^a (\x)\, ,\,\bar{K} \big\}
\\
\qquad\text{where}\qquad
\bar{K} := \int_\Sigma\!  {\rm d}^3x \, K_m{}^a \tilde{E}_a{}^m\,,
\end{multline}
and then eliminating the integrand of $\bar{K}$ using
\begin{equation}
\label{e:trick3}
\begin{aligned}
\bar{K} (\x) &= \frac{1}{\gamma^{3/2}}\Big\{ \frac{\tilde{E}_a{}^m
        \tilde{E}_b{}^n}{\sqrt{\tilde{E}}} \epsilon^{abc} F_{mnc}(\x) 
        \,,\, V \Big\}\, \\
&= \frac1{4\gamma^{5/2}} \epsilon^{mnp} \Big\{ \{ A_m{}^a \, ,\, V \} 
     F_{npa} \, ,\, V \Big\} \, , 
\end{aligned}
\end{equation}
that is, writing it as a nested Poisson bracket.
Inserting these tricks into the Hamiltonian constraint, one
replaces~\eqref{e:clH} with the expression
\begin{multline}
\label{e:H_start}
H[N] = \int_\Sigma\!{\rm d}^3x\, N \epsilon^{mnp} \Tr\Big( F_{mn}
\{ A_p, V\} \\[1ex]
- \frac{1}{2}(1+\gamma^2) 
 \{ A_m, \bar{K} \} \{ A_n, \bar{K}\} \{ A_p, V\} \Big)\,,
\end{multline}
with $\bar{K}$ understood to be eliminated using~\eqref{e:trick3}.
This expression, which now contains only the connection~$A$ and the 
volume~$V$, is the starting point for the construction of the
quantum constraint operator.

In order to quantise the classical Hamiltonian \eqref{e:H_start},
one next elevates all classical objects to quantum operators as
described in the foregoing sections, and replaces the Poisson brackets
in \eqref{e:H_start} by quantum commutators. The resulting
\emph{regulated Hamiltonian} then reduces to a sum over the vertices
$v_\alpha$ of the spin network with lapses $N(v_\alpha)$
\begin{multline}
\label{e:mess1}
\hat{H}[N,\epsilon] = \sum_{\alpha} \,  N(v_{\alpha})\, \epsilon^{mnp}\\[1ex]
\begin{aligned}
 & \times\Tr\bigg\{ \big(h_{\partial{P_{mn}(\epsilon)}} 
  - h_{\partial{P_{mn}(\epsilon)}}^{-1} \big) 
\, h^{-1}_{p}\, \big[ h_{p}, \hat{V}\big]  \\[1ex]
 & - \tfrac{1}{2} (1\! + \!\gamma^2)   h^{-1}_{m}\,\big[ h_{m}, \bar{K} \big] \,
  h_{n}^{-1}\, \big[ h_{n}, \bar{K} \big]\, h_{p}^{-1} \big[h_{p}, \hat{V} \big]\!   \bigg\} \,,
\end{aligned}
\end{multline}
where  $\partial P_{mn}(\epsilon)$ is a small loop attached to the vertex
$v_\alpha$ that must eventually be shrunk to zero. In writing the above
expression, we have furthermore assumed a specific (but, at this point,
not specially preferred) ordering of the operators.

\begin{figure}[t]
\begin{center}
\psfrag{a1}{}
\psfrag{a2}{}
\psfrag{a3}{}
\psfrag{b1}{}
\psfrag{b2}{}
\psfrag{b3}{}
\psfrag{l1}{}
\psfrag{l2}{}
\psfrag{l3}{}
\psfrag{j}{\small $j$}
\psfrag{j1}{\small \!\!$j_1$}
\psfrag{j2}{\small \!$j_2$}
\psfrag{j3}{\small $j_3$}
\psfrag{k1}{\small \!\!$k_1$}
\psfrag{k2}{\small \!\!$k_2$}
\psfrag{k3}{\small $k_3$}
\psfrag{k}{\small $k$}
\psfrag{C}{\small $C$}
\psfrag{V}{}
\vspace{3ex}
\includegraphics[width=\columnwidth]{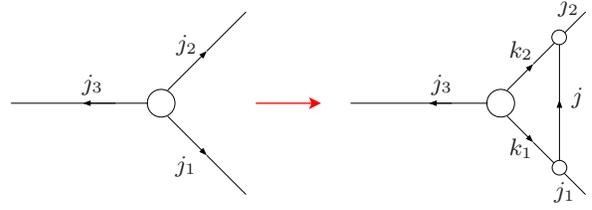}
\end{center}
\caption{Schematic depiction of the action of the Hamiltonian
  constraint on a vertex of a spin network wave function. Two new
  vertices are introduced, and the original vertex is modified. Note
  that in order for this to be true, particular choices have been made
  in the quantisation prescription.\label{f:Hact}}
\end{figure}

Working out the action of~\eqref{e:mess1} on a given spin network wave
function is rather non-trivial, and we are not aware of any concrete
calculations in this regard, other than for very simple special
configurations (see e.g.~\cite{DePietri:1996pj}); to get an idea of
the complications, readers may have a look at a recent analysis of the
volume operator and its spectrum in~\cite{Brunnemann:2004xi}. In
particular, the available calculations focus almost exclusively on the
action of the first term in~\eqref{e:mess1}, whereas the second term
(consisting of multiply nested commutators, cf.~\eqref{e:trick3})
is usually not discussed in any detail. At any rate, this calculation
involves a number of choices in order to fix various ambiguities, 
such as e.g. the ordering ambiguities in both terms in~\eqref{e:mess1}.
An essential ingredient is the action of the operator
$h_{\partial{P_{mn}(\epsilon)}} -
h_{\partial{P_{mn}(\epsilon)}}^{-1}$, which is responsible for the
addition of a plaquette to the spin network. The way in which this
works is depicted (schematically) in figure~\ref{f:Hact}.  The
plaquette is added in a certain SU(2) representation, corresponding to
the representation of the trace in~\eqref{e:mess1}.  This representation 
label~$j$ is arbitrary, and constitutes a quantisation ambiguity
(often called `$m$-ambiguity').

\begin{figure*}[t]
\setcaptionwidth{.9\textwidth}
\begin{center}
\vspace{-3ex}
\psfrag{eps}{$\epsilon$}
\psfrag{H}{$H$}
\vspace{3ex}
\includegraphics*[width=.65\textwidth]{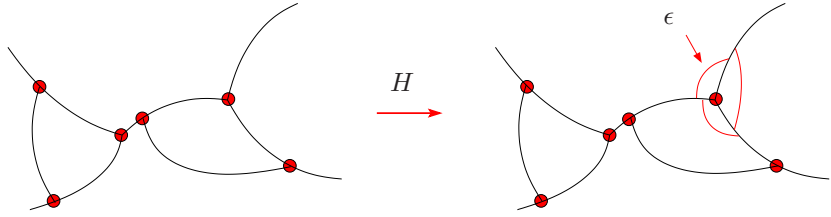}\vspace{-1ex}
\end{center}
\caption{The action of the Hamiltonian constraint is ``ultra-local'',
  in the sense that it acts only in a neighbourhood of ``size''
  $\epsilon$ around a spin network vertex.\label{f:ultralocal}}
\end{figure*}

Having defined the action of the regulated Hamiltonian, the task is
not finished, however, because one must still take the limit $\epsilon
\rightarrow 0$, in which the attached loops are shrunk to zero.  As it
turns out, this limit cannot be taken straightforwardly: due to the
scalar product~\eqref{e:scalar_product} and the non-separability of
$\Hk$ the limiting procedure runs through a sequence of mutually
orthogonal states, and therefore does not converge in $\Hk$. For this
reason, LQG must resort to a weaker notion of limit, either by
defining the limit as a weak limit on a (subspace of the) algebraic
dual of a dense subspace of
$\Hk$~\cite{Lewandowski:1997ba,Ashtekar:2004eh}, or by taking the
limit in the \mbox{weak~$^*$ operator}
topology~\cite{Thiemann:2001yy}.  In the first case the relevant space
(sometimes referred to as the `habitat') is a distribution space which
contains the space $\Hd$ of formally diffeomorphism invariant states
as a subspace, but its precise nature and definition is still a matter
of debate. In the second case, the limit is implemented (in a very
weak sense) on the original kinematical Hilbert space $\Hk$, but that
space will not contain any diffeomorphism invariant states other than
the `vacuum' $\Psi={\bf 1}$. The question of the proper `habitat' on
which to implement the action of the Hamiltonian constraint is thus by
no means conclusively settled.

From a more general point of view, it should be noted that the action
of the Hamiltonian constraint is always `ultralocal': all changes to
the spin network are made in an $\epsilon \rightarrow 0$ neighbourhood
of a given vertex, while the spin network graph is kept
fixed~\mbox{\cite{Smolin:1996fz,Neville:1998eb,Loll:1997iw}}. Pictorially 
speaking, the only action of the (regulated) Hamiltonian is to dress
up the vertices with `spiderwebs', see figure~\ref{f:ultralocal}. More
specifically, it has been argued~\cite{Lewandowski:1997ba} that the
Hamiltonian acts at a particular vertex only by changing the
intertwiners at that vertex. This is in stark contrast to what happens
in lattice field theories. There the action of the Hamiltonian always
links two different existing nodes, the plaquettes are by construction
always spanned between existing nodes, and the continuum limit
involves the lattice as a whole, not only certain sub-plaquettes that
shrink to a vertex. This is also what one would expect on physical
grounds for a theory with non-trivial dynamics.

The attitude often expressed with regard to the ambiguities in the 
construction of the Hamiltonian is that they  correspond to \emph{different} 
physics, and therefore the choice of the correct Hamiltonian is ultimately 
a matter of physics (experiment?), and not mathematics. However, it appears 
unlikely to us that Nature will allow such a great degree of arbitrariness 
at its most fundamental level: in fact, our main point here is that the 
infinitely many ambiguities which killed perturbative quantum gravity, 
are also a problem that other (to wit, non-perturbative) approaches must 
address and solve.~\footnote{The abundance of `consistent' Hamiltonians 
and spin foam models (see below) is sometimes compared to the vacuum 
degeneracy problem of string theory, but the latter concerns
\emph{different solutions} of the \emph{same} theory, as there is no dispute
as to what (perturbative) string theory \emph{is}. However, the concomitant 
lack of predictivity is obviously a problem for both approaches.}

\section{Quantum space-time covariance?}
\label{s:covariance}

Spacetime covariance is a central property of Einstein's theory.
Although the Hamiltonian formulation is not manifestly covariant, full
covariance is still present in the classical theory, albeit in a
hidden form, via the classical (Poisson or Dirac) algebra of
constraints acting on phase space. However, this is not necessarily so
for the quantised theory. As we explained, LQG treats the 
diffeomorphism constraint and the Hamiltonian constraint in
a very different manner. Why and how then should one expect such a
theory to recover full spacetime (as opposed to purely spatial)
covariance?  The crucial issue here is clearly what LQG has to say
about the quantum algebra of constraints.  Unfortunately, to the best
of our knowledge, the `off-shell' calculation of the commutator of two
Hamiltonian constraints in LQG -- with an explicit operatorial
expression as the final result -- has never been fully carried
out. Instead, a survey of the possible terms arising in this
computation has led to the conclusion that the commutator vanishes on
a certain restricted `habitat' of
states~\cite{Thiemann:1996ay,Gambini:1997bc,Lewandowski:1997ba}, and
that therefore the LQG constraint algebra closes without anomalies. By
contrast, we have argued in~\cite{kas_lqg} that this `on shell closure' 
is not sufficient for a full proof of \emph{quantum spacetime covariance},
but that a proper theory of quantum gravity requires a constraint
algebra that closes `off shell', {\it i.e.\/} without prior imposition
of a subset of the constraints. The fallacies that may ensue if one
does not insist on off-shell closure can be illustrated with simple
examples. In our opinion, this requirement may well provide the acid
test on which any proposed theory of canonical quantum gravity will
stand or fail.

While there is general agreement as to what one means when one speaks
of `closure of the constraint algebra' in classical gravity (or any
other classical constrained system~\cite{Henneaux:1992ig}), this 
notion is more subtle 
in the quantised theory. \footnote{For reasons of space, we here restrict  
attention to the bracket between two Hamiltonian constraints, because
the remainder of the algebra involving the kinematical constraints
is relatively straightforward to implement.}
Let us therefore clarify first the various notions
of closure that can arise: we see at least three different
possibilities.  The strongest notion is `off-shell closure' (or
`strong closure'), where one seeks to calculate the commutator of
two Hamiltonians
\begin{equation}\label{HH1}
\big[ \hH [N_1] \, , \, \hH [N_2] \big] = \hO(N_1;N_2)\,.
\end{equation}
Here we assume that the quantum Hamiltonian constraint operator 
\begin{equation}
\hH[N] := \lim_{\epsilon\rightarrow 0} \hH[N,\epsilon] \, ,
\end{equation}
has been rigorously defined as a suitably weak limit, and without 
further restrictions on the states on which (\ref{HH1}) is supposed 
to hold. In writing the above equations, we have thus been (and
will be) cavalier about habitat questions and the precise definition
of the Hamiltonian; see, however
\cite{Lewandowski:1997ba,Gambini:1997bc,kas_lqg} for further details
and critical comments.

Unfortunately, it appears that the goal of determining
$\hO(N_1;N_2)$ as a \emph{bona fide} `off-shell' operator on a suitable
`habitat' of states, and prior to the imposition of any constraints, is 
unattainable within the current framework of LQG. For this reason, LQG must 
resort to weaker notions of closure, by making partial use of the
constraints. More specifically, equation~\eqref{HH1} can be relaxed
substantially by demanding only
\begin{equation}\label{HH2}
\big[ \hH [N_1] \, , \, \hH [N_2] \big]\, |{\cal{X}}\rangle = 0\,,
\end{equation}
but still with the unregulated Hamiltonian constraint $\hH[N]$.
This `weak closure' should hold for all states $|\cal{X}\rangle$ in a 
restricted habitat of states that are `naturally' expected to be 
annihilated by the r.h.s.~of \eqref{HH1}, and that are subject to the 
further requirement that the Hamiltonian can be applied twice without 
leaving the `habitat'. The latter condition is, for instance, met by
the `vertex smooth' states of~\cite{Lewandowski:1997ba}. As shown in
~\cite{Gambini:1997bc,Lewandowski:1997ba}, the commutator of two 
Hamiltonians indeed vanishes on this `habitat', and one is therefore 
led to conclude that the full constraint algebra closes `without anomalies'. 

The same conclusion was already arrived at in an earlier computation
of the constraint algebra in~\cite{Thiemann:1996aw,Thiemann:1996ay},
which was done from a different perspective (no `habitats'), and makes
essential use of the space of diffeomorphism invariant states $\Hd$, the 
`natural' kernel of the r.h.s.~of \eqref{HH1}. Here the idea is to
verify that~\cite{Thiemann:1996aw,Thiemann:1996ay}
\begin{equation}
\label{HH3}
\lim_{\substack{\epsilon_1\rightarrow 0\\\epsilon_2\rightarrow 0}}
\langle {\cal X}\, | \, \big[ \hH[N_1,\epsilon_1] \, , \, \hH [N_2,\epsilon_2] \big]
\,\Psi\rangle =0\,,
\end{equation}
for all $|{\cal X}\rangle\in\Hd$, and for all~$|\Psi\rangle$ in the
space of finite linear combinations of spin network states. As for the
Hamiltonian itself, letting $\epsilon_{1,2}\rightarrow 0$ in this
expression produces an uncountable sequence of mutually orthogonal
states w.r.t.~the scalar product
\eqref{e:scalar_product}. Consequently, the limit again does not exist
in the usual sense, but only as a \wsl.  The `diffeomorphism
covariance' of the Hamiltonian is essential for this result.
Let us stress that \eqref{HH2} and \eqref{HH3} are by no means the
same: in \eqref{HH2} one uses the unregulated Hamiltonian (where the
limit $\epsilon\rightarrow 0$ has already been taken), whereas the
calculation of the commutator in~\eqref{HH3} takes place inside~$\Hk$,
and the limit $\epsilon\rightarrow 0$ is taken only \emph{after}
computing the commutator of two regulated Hamiltonians. These two
operations (taking the limit $\epsilon\rightarrow 0$, and calculating
the commutator) need not commute. Because with both~\eqref{HH2}
and~\eqref{HH3} one forgoes the aim of finding an operatorial
expression for the commutator $\big[\hat{H}[N_1], \hat{H}[N_2] \big]$,
making partial use of the constraints, we say (in a partly
supergravity inspired terminology) that the algebra closes `on-shell'.

Although on-shell closure may perhaps look like a sufficient condition
on the quantum Hamiltonian constraint, it is easy to see, at the level
of simple examples, that this is not true. Consider, for instance, the
Hamiltonian constraint of bosonic string theory, and consider
modifying it by multiplying it with an operator which commutes with
all Virasoro generators. There are many such operators in string
theory, for instance the mass-squared operator (minus an arbitrary
integer). In this way, we arrive at a realisation of the constraint
operators which is very similar to the one used in LQG: the algebra of
spatial diffeomorphisms is realised via a (projective) unitary
representation, and the Hamiltonian constraint transforms covariantly
(the extra factor does not matter, because it commutes with all
constraints). In a first step, one can restrict attention to the
subspace of states annihilated by the diffeomorphism constraint, the
analog of the space~$\Hd$. Imposing now the new Hamiltonian constraint
(the one multiplied with the Casimir) on this subspace would 
produce a `non-standard' spectrum by allowing extra diffeomorphism
invariant states of a certain prescribed mass. The algebra would also
still close on-shell, i.e.~on the `habitat' of states annihilated by
the diffeomorphism constraint. The point here is not so much whether
this new spectrum is `right' or `wrong', but rather that in allowing
such modifications which are compatible with on-shell closure of the
constraint algebra, we introduce an infinite ambiguity and
arbitrariness into the definition of the physical states. In other
words, if we only demand on-shell closure as in LQG, there is no way
of telling whether or not the vanishing of a commutator is merely
accidental, that is, not really due to the diffeomorphism invariance
of the state, but caused by some other circumstance.

By weakening the requirements on the constraint algebra and by no
longer insisting on off-shell closure, crucial information gets
lost. This loss of information is reflected in the ambiguities
inherent in the construction of the LQG Hamiltonian.  It is quite
possible that the LQG Hamiltonian admits many further modifications on
top of the ones we have already discussed, for which the commutator
continues to vanish on a suitably restricted habitat of states --- in
which case neither \eqref{HH2} nor \eqref{HH3} would amount to much of
a consistency test.

\section{Canonical gravity and spin foams}
\label{s:net2foam}

\begin{figure}[t]
\begin{center}
\psfrag{j}{$j$}
\includegraphics*[width=.7\columnwidth]{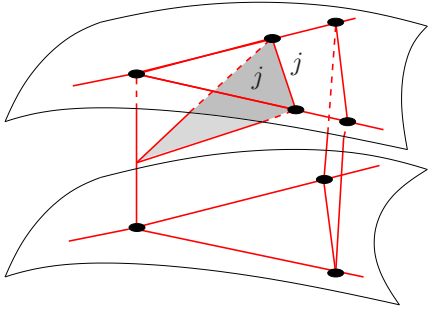}
\end{center}
\caption{From spin networks to spin foams, in (2+1) dimensions. The
  Hamiltonian constraint has created one new edge and two new
  vertices. The associated surface inherits the label~$j$ of the edge
  which is located on the initial or (in this case) final space-like
  surface.\label{f:tofoams}}
\end{figure}

Attempts to overcome the difficulties with the Hamiltonian constraint
have led to another development, \emph{spin foam
models}~\cite{Reisenberger:1994aw,Reisenberger:1997pu,Baez:1997zt}.
These were originally proposed as space-time versions of spin
networks, to wit, evolutions of spin networks in `time', but have
since developed into a class of models of their own, disconnected from 
the canonical formalism. Mathematically, spin foam models represent a
generalisation of spin networks, in the sense that group theoretical
objects (holonomies, representations, intertwiners, etc.) are attached
not only to vertices and edges (links), but also to higher dimensional
faces in a simplicial decomposition of space-time.

The relation between spin foam models and the canonical formalism is
based on a few general features of the action of the Hamiltonian
constraint operator on a spin network (for a review on the connection,
see~\cite{DePietri:1999cp}). As we have discussed above, the
Hamiltonian constraint acts, schematically, by adding a small
plaquette close to an existing vertex of the spin network (as in
figure~\ref{f:Hact}). In terms of a space-time picture, we see that 
the edges of the spin network sweep out surfaces, and the Hamiltonian 
constraint generates new surfaces, as in figure~\ref{f:tofoams}; 
but note that this graphical representation does not capture the 
details of how the action of the Hamiltonian affects the intertwiners 
at the vertices. Instead of associating spin labels to the
edges of the spin network, one now associates the spin labels to the
surfaces, in such a way that the label of the surface is determined by
the label of the edge which lies in either the initial or final
surface.

In analogy with proper-time transition amplitudes for a relativistic
particle, it is tempting to define the transition amplitude between an
initial spin network state and a final one as
\begin{multline}
\label{e:projector}
Z_T
 := \langle \psi_f | \exp \Big(i\!\int_0^T\! {\rm d}t\,H\Big)\, |\psi_i\rangle\\[1ex]
= \sum_{n=0}^\infty \frac{(i\,T)^n}{n!}
\int\!{\rm d}\psi_1\ldots {\rm d}\psi_n\,
\langle \psi_f | H | \psi_1\rangle\\[1ex]
\times \langle \psi_1 | H | \psi_2\rangle 
\cdots \langle \psi_n | H | \psi_i\rangle\,,
\end{multline}
where we have repeatedly inserted resolutions of unity. A (somewhat
heuristic) derivation of the above formula can be given by starting
from a formal path integral \cite{Reisenberger:1997pu}, which, after
gauge fixing and choice of a global time coordinate $T$, and with
appropriate boundary conditions, can be argued to reduce to the above
expression. There are many questions one could ask about the physical
meaning of this expression, but one important property is that (just
as with the relativistic particle), the transition amplitude will
project onto physical states (formally, this projection is effected in
the original path integral by integrating over the lapse function
multiplying the Hamiltonian density). One might thus
consider~\eqref{e:projector} as a way of defining a physical inner product.

Because path integrals with oscillatory measures are notoriously difficult 
to handle, one might wonder at this point whether to apply a formal
Wick rotation to~\eqref{e:projector}, replacing the Feynman weight 
with a Boltzmann weight, as is usually done in Euclidean quantum field 
theory. This is also what is suggested by the explicit formulae in 
\cite{Reisenberger:1997pu}, where $i$ in (\ref{e:projector}) is replaced 
by $(-1)$. However, this issue is much more subtle here than in ordinary 
(flat space) quantum field theory. First of all, the distinction between 
a Euclidean (Riemannian) and a Lorentzian (pseudo-Riemannian) manifold 
obviously requires the introduction of a metric of appropriate
signature. However, spin foam models, having their roots in (background 
independent) LQG, do not come with a metric, and thus the terminology 
is to some extent up to the beholder. To avoid confusion, let us state 
clearly that our use of the words `Euclidean' and `Lorentzian' here
always refers to the use of oscillatory weights~$e^{i S_E}$ and $e^{i S_L}$,
respectively, where the actions~$S_E$ and $S_L$ are the respective
actions for Riemannian resp.~pseudo-Riemannian metrics. The term
`Wick rotated', on the other hand, refers to the replacement of the
oscillatory weight $e^{iS}$ by the exponential weight $e^{-S}$,
with either $S=S_E$ or $S=S_L$. However, in making use of this terminology, 
one should always remember that there is no Osterwalder-Schrader type 
reconstruction theorem in quantum gravity, and therefore any procedure
(or `derivation') remains formal. Unlike the standard 
Euclidean path integral~\cite{Gibbons:1976ue,hawk3}, the spin foam
models to be discussed below are generally interpreted to correspond 
to path integrals \emph{with oscillatory weights} $e^{iS}$, but come
in both Euclidean and Lorentzian variants (corresponding to the 
groups SO(4) and SO(1,3), respectively). This is true even if the
state sums involve only \emph{real} quantities ($nj$-symbols, edge
amplitudes, etc.), cf.~the discussion after~\eqref{e:6jlarge}.

\begin{figure}[t]
\begin{center}
\psfrag{psif}{$|\psi_f\rangle$}
\psfrag{psii}{$|\psi_i\rangle$}
\psfrag{j1}{\smaller $j_1$}
\psfrag{j2}{\smaller $j_2$}
\psfrag{j3}{\smaller $j_3$}
\psfrag{j4}{\smaller $j_4$}
\psfrag{j5}{\smaller $j_5$}
\psfrag{j6}{\smaller $j_6$}
\includegraphics*[width=\columnwidth]{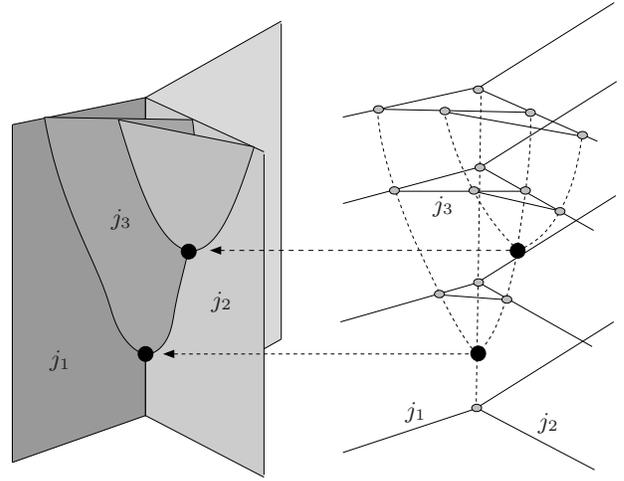}
\end{center}
\caption{A spin foam (left) together with its spin network evolution
  (right) in (2+1) dimensions. Spin foam nodes correspond to the
  places where the Hamiltonian constraint in the spin network acts
  non-trivially (black dots).  Spin foam edges correspond to evolved
  spin network nodes (grey dots), and spin foam faces correspond to
  spin network edges.  The spin labels of the faces are inherited from
  the spin labels of spin network edges. If all spin network nodes are
  three-valent, the spin foam nodes sit at the intersection of six
  faces, and the dual triangulation consists of
  tetrahedrons.\label{f:spinfoams_vs_regge}}
\end{figure}

The building blocks~$\langle \psi_{k} | H |\psi_{l}\rangle$ in the
transition amplitude~\eqref{e:projector} correspond to elementary spin
network transition amplitudes, as in figure~\ref{f:tofoams}. For a
given value of~$n$, i.e.~a given number of time slices, we should thus
consider objects of the type
\begin{equation} 
\label{e:state_sum}
Z_{\psi_1,\ldots,\psi_n} = \langle \psi_f | H | \psi_1\rangle
\langle \psi_1 | H | \psi_2\rangle 
\cdots \langle \psi_n | H | \psi_i\rangle\,.
\end{equation}
Each of the building blocks depends only on the values of the spins at
the spin network edges and the various intertwiners in the spin
network state. The points where the Hamiltonian constraint acts
non-trivially get associated to spin foam vertices; see
figure~\ref{f:spinfoams_vs_regge}. Instead of working
out~\eqref{e:state_sum} directly from the action of the Hamiltonian
constraint, one could therefore also define the amplitude directly in
terms of sums over expressions which depend on the various spins
meeting at the spin foam nodes. In this way, one arrives at so-called
\emph{state sum} models, which we will describe in the following section.

\begin{figure*}[t]
\setcaptionwidth{.9\textwidth}
\begin{center}
\includegraphics*[width=.65\textwidth]{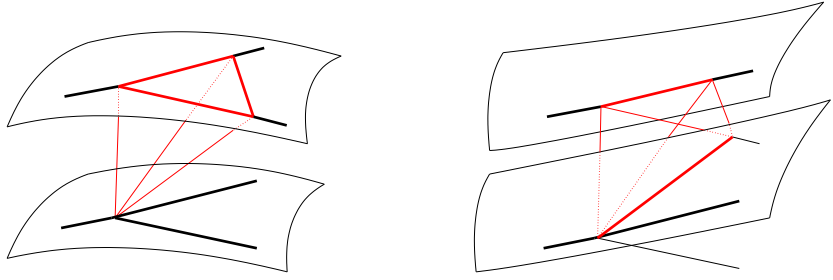}
\end{center}
\caption{The Hamiltonian constraint induces a $1\!\rightarrow\! 3$~move in
  the spin foam formalism (figure on the left). However, by slicing
  space-time in a different way, one can equivalently interpret this
  part of the spin foam as containing a $2\!\rightarrow\! 2$~move (figure
  on the right). This argument suggests that the ultra-local
  Hamiltonian may not be sufficient to achieve space-time covariance.
  For clarity, the network edges which lie in one of the spatial slices 
  have been drawn as thick lines.\label{f:moves2}}
\end{figure*}

A problematic issue in the relation between spin foams and the
canonical formalism comes from covariance requirements. While
tetrahedral symmetry (or the generalisation thereof in four
dimensions) is natural in the spin foam picture, the action of the
Hamiltonian constraint, depicted in figure~\ref{f:tofoams}, does not
reflect this symmetry. The Hamiltonian constraint only leads to
so-called $1\!\rightarrow\! 3$~moves, in which a single vertex in the
initial spin network is mapped to three vertices in the final spin
network. In the spin foam picture, the restriction to only these moves
seems to be in conflict with the idea that the slicing of space-time
into a space+time decomposition can be chosen arbitrarily. For
space-time covariance, one expects $2\!\rightarrow\! 2$ and $0\!
\rightarrow\!  4$ moves (and their time-reversed partners) as well,
see figure~\ref{f:moves2}. These considerations show that there is
no unique path from canonical gravity to spin foam models, and thus no
unique model either (even if there was a unique canonical Hamiltonian). 

It has been argued~\cite{Reisenberger:1997pu} that these missing moves
can be obtained from the Hamiltonian formalism by a suitable choice of
operator ordering. In section~\ref{s:constraints} we have used an
ordering, symbolically denoted by~$FEE$, in which the Hamiltonian
first opens up a spin network and subsequently glues in a plaquette.
If one chooses the ordering to be~$EEF$, then the inverse densitised
vielbeine can open the plaquette, thereby potentially inducing a
$2\!\rightarrow\! 2$~or $0\!\rightarrow\! 4$~move. However, 
ref.~\cite{Thiemann:1996aw} has argued strongly against this operator 
ordering, claiming that in such a form the Hamiltonian operator 
cannot even be densely defined. In addition, 
the derivation sketched here is rather symbolic and hampered
by the complexity of the Hamiltonian constraint~\cite{Rovelli:1998dx}. 
Hence, to summarise: for (3+1) gravity a decisive proof of the connection 
between spin foam models and the full Einstein theory and its canonical 
formulation appears to be lacking, and it is by no means excluded that 
such a link does not even exist.

\section{Spin foam models: some basic features}

In view of the discussion above, it is thus perhaps best to view spin
foam models as models in their own right, and, in fact, as a novel way of
defining a (regularised) path integral in quantum gravity. Even without 
a clear-cut link to the canonical spin network quantisation programme,
it is conceivable that spin foam models can be constructed which possess
a proper semi-classical limit in which the relation to classical gravitational 
physics becomes clear. For this reason, it has even been suggested that 
spin foam models may provide a possible `way out' if the difficulties with 
the conventional Hamiltonian approach should really prove insurmountable.

The simplest context in which to study state sum models is 
(2+1) gravity, because it is a topological (`BF-type') theory, that is, 
without local degrees of freedom, which can be solved exactly (see
e.g.~\mbox{\cite{Deser:1983tn,Witten:1988hc,Ashtekar:1989qd}}
and~\cite{Noui:2004iy} for a more recent analysis of the model within
the spin foam picture).  The most general expression for a state sum
in (2+1) dimensions takes, for a given spin foam~$\phi$, the form
\begin{multline}
\label{e:statesum}
Z_\phi = 
\sum_{\text{spins $\{j\}$}\strut}\;
\prod_{f,e,v\strut}   A_f(\{j\})\, A_e(\{j\})\, A_v(\{j\})\,,
\end{multline}
where $f,e,v$ denote the faces, edges and vertices respectively. The
amplitudes depend on all these sub-simplices, and are denoted
by~$A_f$, $A_e$ and $A_v$ respectively. There are many choices which
one can make for these amplitudes. In three Euclidean dimensions,
space-time covariance demands that the contribution to the partition
sum has tetrahedral symmetry in the six spins of the faces which meet
at a node (here we assume a `minimal' spin foam; models with more
faces intersecting at an edge are of course possible). 

Now, a model of this type has been known for a long time: it is the
Ponzano-Regge model for 3d gravity, which implements the above
principles by defining the partition sum
\begin{multline}
\label{e:PR}
Z^{\text{PR}}_\phi 
=\!\!\!\! \sum_{\text{spins $\{j_i\}$}}\; \prod_{\text{faces $f$}} (2\,j_f + 1)
  \prod_{\text{vertices $v$}} \raisebox{-3ex}{\psfrag{j1}{\smaller\smaller\smaller\smaller $j_1$}
\psfrag{j2}{\smaller\smaller\smaller\smaller $j_3$}
\psfrag{j3}{\smaller\smaller\smaller\smaller $j_4$}
\psfrag{j4}{\smaller\smaller\smaller\smaller $j_6$}
\psfrag{j5}{\smaller\smaller\smaller\smaller $j_5$}
\psfrag{j6}{\smaller\smaller\smaller\smaller $j_2$}
\includegraphics[width=8ex]{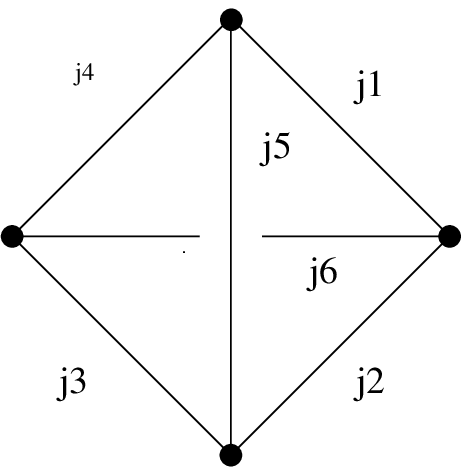}}
\end{multline}
The graphical notation denotes the Wigner $6j$ symbol, defined in
terms of four contracted Clebsch-Gordan coefficients as
\begin{equation}
\label{e:6j}
\{ 6j \}
\sim\!\!\!\!\!\! \sum_{m_1,\ldots,m_6}\!\!\!
C{}^{j_1}_{m_1}{}^{j_2}_{m_2}{}^{j_3}_{m_3} 
C{}^{j_5}_{m_5}{}^{j_6}_{m_6}{}^{j_1}_{m_1} 
C{}^{j_6}_{m_6}{}^{j_4}_{m_4}{}^{j_2}_{m_2}
C{}^{j_4}_{m_4}{}^{j_5}_{m_5}{}^{j_3}_{m_3}\,.
\end{equation}
For SU(2) representations, the sum over spins in the Ponzano-Regge
state sum~\eqref{e:PR} requires that one divides by an infinite factor
in order to ensure convergence (more on finiteness properties below)
and independence of the triangulation.  The tetrahedron appearing
in~\eqref{e:PR} in fact has a direct geometrical interpretation as the
object dual to the spin foam vertex. The dual tetrahedron can then
also be seen as an elementary simplex in the triangulation of the
manifold. Three-dimensional state sums with boundaries, appropriate
for the calculation of transition amplitudes between two-dimensional
spin networks, have been studied in~\cite{Karowski:1991ke}.

When one tries to formulate spin foam models in four dimensions, the
first issue one has to deal with is the choice of the representation
labels on the spin foam faces. From the point of view of the canonical
formalism it would seem natural to again use SU(2) representations, as
these are used to label the edges of a spin network in three spatial
dimensions, whose evolution produces the faces (2-simplices) of the
spin foam.  However, this is not what is usually done. Instead, the
faces of the spin foam are supposed to carry representations of SO(4)
$\approx$ SO(3) $\times$ SO(3) [or SO(1,3) $\approx$
SL(2,$\mathbb{C}$) for Lorentzian space-times].  The corresponding
models in four dimensions are purely topological theories, the
so-called ``$BF$ models'', where~$F(A)$ is a field strength, and~$B$
the Lagrange multiplier two-form field whose variation
enforces~$F(A)=0$. Up to this point, the model is analogous to gravity
in (2+1) dimensions, except that the relevant gauge group is now SO(4)
[or SO(1,3)]. However, in order to recover general relativity and to
re-introduce local (propagating) degrees of freedom into the theory,
one must impose a constraint on~$B$.

Classically, this constraint says that $B$ is a `bi-vector', that is
$B^{ab} = e^a\wedge e^b$. The quantum mechanical analog of this
constraint amounts to a restriction to a particular set of
representations of $\text{SO}(4) = \text{SU}(2)\otimes \text{SU}(2)$,
namely those where the spins of the two factors are equal, the
so-called \emph{balanced representations}, denoted by $(j,j)$ (for
$j=\frac12,1,\frac32, \dots$). Imposing this restriction on the state
sum leads to a class of models first proposed by Barrett \&
Crane~\cite{Barrett:1997gw,Freidel:1998pt}.  In these models the
vertex amplitudes are given by combining the 10 spins of the faces
which meet at a vertex, as well as possibly further `virtual' spins
associated to the vertices themselves, using an expression built from
contracted Clebsch-Gordan coefficients. For instance, by introducing
an extra `virtual' spin~$i_k$ associated to each edge where four faces
meet, one can construct an intertwiner between the four spins by means
of the following expression
\begin{equation}\label{e:I}
I{}^{j_1}_{m_1}{}^{\cdots}_{\cdots}{}^{j_4}_{m_4}{}^{;i_k} = 
\sum_{m_k} C{}^{j_1}_{m_1}{}^{j_2}_{m_2}{}^{i_k}_{m_k}
C{}^{j_3}_{m_3}{}^{j_4}_{m_4}{}^{i_k}_{m_k}\,.
\end{equation}
However, this prescription is not unique as we can choose between
three different `channels' (here taken to be $12 \leftrightarrow 34$);
this ambiguity can be fixed by imposing symmetry, see
below. Evidently, the number of channels and virtual spins increases
rapidly with the valence of the vertex. For the above four-vertex,
this prescription results in a state sum\footnote{There is now no
longer such a clear relation of the graphical object in~\eqref{e:BC}
to the dual of the spin foam vertex: faces and edges of the spin foam
map to faces and tetrahedrons of the dual in four dimensions,
respectively, but these are nevertheless represented with edges and
vertices in the figure in~\eqref{e:BC}.}
\begin{multline}
\label{e:BC}
Z^{\{i_k\}}_\phi 
= \sum_{\text{spins $\{j_i\}$}}\; \prod_{\text{faces $f$}}\; \prod_{\text{edges $e$}} A_f(\{j\})\;
  A_e(\{j\})\\[-1ex]
\times  \prod_{\text{vertices $v$}}\; \raisebox{-6ex}{\psfrag{j1}{\smaller\smaller\smaller\smaller $j_1$}
\psfrag{j2}{\smaller\smaller\smaller\smaller $j_2$}
\psfrag{j3}{\smaller\smaller\smaller\smaller $j_{0}$}
\psfrag{j4}{\smaller\smaller\smaller\smaller $j_6$}
\psfrag{j5}{\smaller\smaller\smaller\smaller $j_5$}
\psfrag{j6}{\smaller\smaller\smaller\smaller $j_4$}
\psfrag{j7}{\smaller\smaller\smaller\smaller $j_7$}
\psfrag{j8}{\smaller\smaller\smaller\smaller $j_8$}
\psfrag{j9}{\smaller\smaller\smaller\smaller $j_9$}
\psfrag{j10}{\smaller\smaller\smaller\smaller $j_{3}$}
\psfrag{i1}{\smaller\smaller\smaller\smaller $i_1$}
\psfrag{i2}{\smaller\smaller\smaller\smaller $i_2$}
\psfrag{i3}{\smaller\smaller\smaller\smaller $i_3$}
\psfrag{i4}{\smaller\smaller\smaller\smaller $i_4$}
\psfrag{i5}{\smaller\smaller\smaller\smaller $i_5$}
\includegraphics[width=14ex]{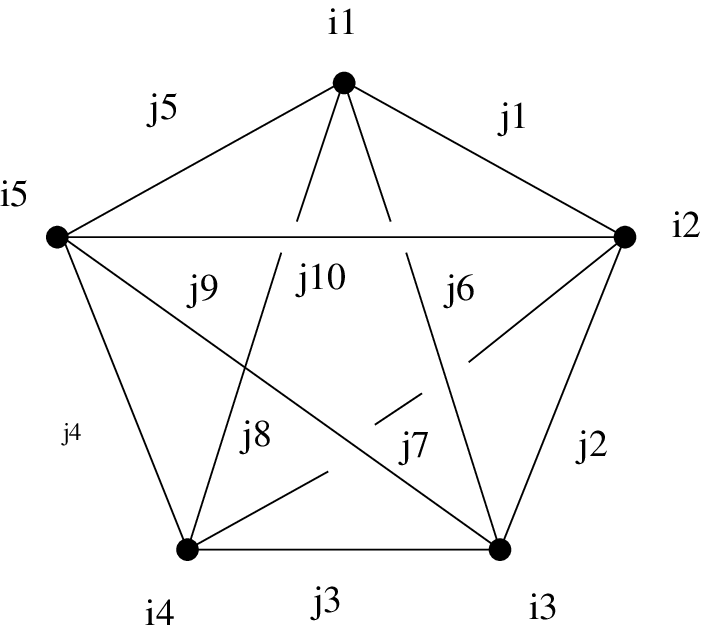}}\,,
\end{multline}
where the spins~$j$ denote spin labels of balanced
representations~$(j,j)$ (as we already mentioned, without this
restriction, the model above corresponds to the topological BF
model~\mbox{\cite{Ooguri:1992eb,Crane:1993if,Crane:1994ji}}).  The
precise factor corresponding to the pentagon (or ``$15j$'' symbol) in
this formula is explicitly obtained by multiplying the
factors~\eqref{e:I} (actually, one for each SO(3) factor in SO(4)), and
contracting (summing) over the labels $m_i$,
\begin{multline}
\{15j\} = \sum_{m_i} 
   I{}^{j_1}_{m_1}{}^{j_4}_{m_4}{}^{j_9}_{m_9}{}^{j_5}_{m_5}{}^{;i_1}
   I{}^{j_1}_{m_1}{}^{j_2}_{m_2}{}^{j_7}_{m_7}{}^{j_3}_{m_3}{}^{;i_2}\\[1ex]
 \times
   I{}^{j_4}_{m_4}{}^{j_2}_{m_2}{}^{j_8}_{m_8}{}^{j_0}_{m_0}{}^{;i_3}
   I{}^{j_9}_{m_9}{}^{j_7}_{m_7}{}^{j_0}_{m_0}{}^{j_6}_{m_6}{}^{;i_4}
   I{}^{j_5}_{m_5}{}^{j_3}_{m_3}{}^{j_8}_{m_8}{}^{j_6}_{m_6}{}^{;i_5}\,.
\end{multline}
There are various ways in which one can make~\eqref{e:BC} independent
of the spins~$i_k$ associated to the edges. One way is to simply sum
over these spins.  This leads to the so-called ``$15j$ BC model'',
\begin{multline}
Z^{\text{$15j$}}_\phi = 
    \sum_{\text{spins $\{j_i, i_k\}$}}\; \prod_{\text{faces $f$}}\; \prod_{\text{edges $e$}} A_f(\{j\})\;
  A_e(\{j\})\\[1ex]
\times  \prod_{\text{vertices $v$}}\; \big\{ 15j \big\}\,.
\end{multline}
An alternative way to achieve independence of the edge intertwiner
spins is to include a sum over the~$i_k$ in the definition of the
vertex amplitude. These models are known as ``$10j$ BC models'',
\begin{multline}
\label{e:Z10j}
Z^{\text{$10j$}}_\phi = 
   \sum_{\text{spins $\{j_i\}$}}\; \prod_{\text{faces $f$}}\; \prod_{\text{edges $e$}} A_f(\{j\})\;
  A_e(\{j\})\\[1ex]
\times  \prod_{\text{vertices $v$}}\; \sum_{\text{spins $\{i_k\}$}}
   f(\{i_k\}) \; \big\{ 15j \big\}\,,
\end{multline}
labelled by an arbitrary function~$f(\{i_k\})$ of the intertwiner
spins. Only for the special choice~\cite{Barrett:1997gw} 
\begin{equation}
\label{e:2jp1}
f(\{i_k\}) = \prod_{k=1}^5 (2 i_k + 1)
\end{equation}
does the vertex amplitude have simplicial
symmetry~\cite{Reisenberger:1998bn}, i.e.~is invariant under the
symmetries of the pentagon~\eqref{e:BC} (where the pentagon really
represents a 4-simplex).\footnote{There is an interesting way to
express combinatorial objects such as the $10j$ symbol in terms of
integrals over group manifolds, which goes under the name of `group
field theory' (see e.g.~\cite{Freidel:2005qe}), and which also allows
an interpretation in terms of `Feynman diagrams'. The relation between
spin foams and group field theory is potentially useful to evaluate
state sums because the corresponding integrals can be evaluated using
stationary phase methods. We will, however, not comment on this
development any further since there is (under certain assumptions) a
one-to-one map between spin foam models and group field theory
models.}

While the choice~(\ref{e:Z10j},\ref{e:2jp1}) for the vertex
amplitude~$A_v(\{j\})$ is thus preferred from the point of view of
covariance, there are still potentially many different choices for the
face and edge amplitudes~$A_f(\{j\})$ and $A_e(\{j\})$. Different
choices may lead to state sums with widely varying properties. The
dependence of state sums on the face and edge amplitudes is clearly
illustrated by e.g.~the numerical comparison of various models presented 
in~\cite{Baez:2002aw}. A natural and obvious restriction on the possible 
amplitudes is that the models should yield the correct classical limit 
-- to wit, Einstein's equations -- in the large~$j$ limit, corresponding 
to the infrared (see also the discussion in the following section). 
Therefore, any function of the face spins which satisfies 
the pentagon symmetries and is such that the state sum has appropriate 
behaviour in the large~$j$ limit is a priori allowed. Furthermore,
the number of possible amplitudes, and thus of possible models, grows
rapidly if one allows for more general valences of the vertices. In
the literature, the neglect of higher-valence vertices is often
justified by invoking the fact that the valence~$\leq 4$ spin network
wave functions in the Hamiltonian formulation constitute a
super\-selection sector in $\Hk$ (because the `spiderwebs' in
figure~\ref{f:ultralocal} do not introduce higher valences). However,
we find this argument unconvincing because $(i)$~the precise relation
between the Hamiltonian and the spin foam formulation remains unclear,
and $(ii)$~physical arguments based on ultralocality (cf.~our
discussion at the end of section~\ref{s:net2foam}) suggest that more
general moves (hence, valences) should be allowed.

Let us also mention that, as an alternative to the Euclidean spin foam
models, one can try to set up \emph{Lorentzian spin foam models}, as
has been done in~\mbox{\cite{Barrett:1999qw,Perez:2000ec}}. In this
case, the (compact) group SO(4) is replaced by the \emph{non-compact}
Lorentz group SO(1,3) [or $\text{SL}(2,\mathbb{C})$]. Recall that in
both cases we deal with oscillatory weights, not with a weight
appropriate for a Wick rotated model. It appears unlikely that there
is any relation between the Lorentzian models and the Euclidean ones. 
Furthermore, the analysis of the corresponding Lorentzian state sums is
much more complicated due to the fact that the relevant (i.e.~unitary)
representations are now infinite-dimensional.

\section{Spin foams and discrete gravity}

To clarify the relation between spin foam models and earlier attempts
to define a discretised path integral in quantum gravity, we 
recall that the latter can be roughly divided into two classes, namely:
\begin{itemize}
\item \emph{Quantum Regge Calculus} (see e.g.~\cite{Williams:1991cd}), 
where one approximates space-time by a triangulation consisting of a 
fixed number of simplices, and integrates over all edge lengths, 
keeping the `shape' of the triangulation fixed;
\item \emph{Dynamical Triangulations}~~ (see
e.g.~\cite{Boulatov:1986jd,Billoire:1986ab,Ambjorn:1985dn}), where the
simplices are assigned fixed edge lengths, and one sums instead over
different triangulations, but keeping the number of simplices fixed
(thus changing only the `shape', but not the `volume' of the
triangulation).
\end{itemize}
Both approaches are usually based on a positive signature (Euclidean)
metric, where the Boltzmann factor is derived from, or at least
motivated by, some discrete approximation to the Einstein-Hilbert
action, possibly with a cosmological constant (but
see~\mbox{\cite{Ambjorn:2001cv,Ambjorn:2004qm}} for some recent
progress with a Wick-rotated `Lorentzian' dynamical triangulation
approach which introduces and exploits a notion of causality on the
space-time lattice).  In both approaches, the ultimate aim is then to
recover continuum space-time via a refinement limit in which the
number of simplices is sent to infinity. Establishing the existence of
such a limit is a notoriously difficult problem that has not been
solved for four-dimensional gravity. In fact, for quantum Regge models
in two dimensions such a continuum limit does not seem to agree with
known continuum
results~\cite{Bock:1994mq,Holm:1994un,Ambjorn:1997ub,Rolf:1998ja} (see
however~\cite{Hamber:1997ut}).

From the point of view of the above classification, spin foam models 
belong to the first, `quantum Regge', type, as one sums over all spins
for a given spin foam, but does not add, remove or replace edges, faces 
or vertices, at least not in the first step. Indeed, for the spin foams 
discussed in the foregoing section, we have so far focused on the 
partition sum for a \mbox{\emph{single}} given spin foam. An obvious question 
then concerns the next step, or more specifically the question how spin 
foam models can recover (or even only define) a continuum limit. The
canonical setup, where one sums over all spin network states in 
expressions like~\eqref{e:projector}, would suggest that one should
sum over all foams,
\begin{equation}
Z^\text{total} = \sum_{\text{foams $\phi$}}\; w_\phi\, Z_{\phi}\,,
\end{equation}
where $Z_\phi$ denotes the partition function for a given spin foam
$\phi$, and where we have allowed for the possibility of a non-trivial
weight $w_\phi$ depending only on the topological structure (`shape') of 
the foam. The reason for this sum would be to achieve formal independence
of the triangulations. In a certain sense this would mimic the
dynamical triangulation approach (except that one now would also sum
over foams with a different number of simplices and different edge
lengths), and thus turn the model into a hybrid version of the above
approaches.  However, this prescription is far from universally
accepted, and several other ideas on how to extract classical,
continuum physics from the partition sum~$Z_\phi$ have been proposed.

One obvious alternative is to \emph{not} sum over all foams, but
instead look for a refinement with an increasing number of 
cells,~\footnote{But note that, formally, the sum over all foams can 
also be thought of as a refinement limit if one includes zero spin 
representations (meaning no edge) in the refinement limit.} 
\begin{equation}
\label{e:Zrefined}
Z^\infty = \lim_{\text{\# cells}\rightarrow\infty} Z_\phi\,.
\end{equation}
The key issue is then to ensure that the final result does not depend
on the way in which the triangulations are performed and refined (this
is a crucial step which one should understand in order to interpret
results on single-simplex spin foams like those
of~\cite{Rovelli:2005yj,Speziale:2005ma}). The refinement limit is
motivated by the fact that it does appear to work in three space-time
dimensions, where (allowing for some `renormalisation') one can establish 
triangulation independence~\cite{Ooguri:1991ni}. Furthermore, 
for large spins, the $6j$ symbol which appears in the Ponzano-Regge model 
approximates the Feynman weight for Regge gravity~\cite{ponzano,roberts1}. 
More precisely, when all six spins are simultaneously taken large,
\begin{equation}
\label{e:6jlarge}
\{ 6j\} \sim \left( e^{iS_{\text{Regge}}(\{j\}) + \frac{i\pi}{4}} + 
e^{-iS_{\text{Regge}}(\{j\}) - \frac{i\pi}4} \right) .
\end{equation}
Here~$S_{\text{Regge}}(\{j\})$ is the Regge action of a tetrahedron,
given by
\begin{equation}
S_{\text{Regge}}(\{j\}) = \sum_{i=1}^6 j_i\, \theta_i\,,
\end{equation}
where $\theta_i$ is the dihedral angle between the two surfaces
meeting at the $i$th edge. Related results in four dimensions are
discussed in~\cite{Barrett:1998gs} and, using group field theory
methods, in~\cite{Freidel:1999jf}. We emphasise once more that this by
no means singles out the~$6j$~symbol as the unique vertex amplitude:
we can still multiply it by any function of the six spins which
asymptotes to one for large spins.

The $6j$ symbol is of course real, which explains the presence of a
cosine instead of a complex oscillatory weight on the right-hand side
of~\eqref{e:6jlarge}. Indeed, it seems rather curious that, while the
left-hand side of~\eqref{e:6jlarge} arises from an expression
resembling a Boltzmann sum, the right-hand side contains oscillatory
factors which suggest a path integral with oscillatory weights.  In
view of our remarks in section~\ref{s:net2foam}, and in order to make
the relation to Regge gravity somewhat more precise, one must
therefore argue either that a proper path integral in gravity produces
both terms, or otherwise that one can get rid of one of the terms by
some other mechanisms. The first possibility appears to be realised in
(2+1) gravity, because one can cast the gravitational action into
Chern Simons form $S = \int R\wedge e$, in which case a sum over
orientations of the dreibein would lead to terms with both signs in
the exponent. Unfortunately, this argument does not extend to four
dimensions, where the gravitational action $S = \int R\wedge e \wedge
e$ depends quadratically on the vierbein.  For this reason, it has
instead been suggested that one of the two oscillatory terms
disappears for all physical correlation
functions~\cite{Rovelli:2005yj}.

The vertex amplitudes represented by the~$6j$ or $10j$ symbols only
form part of the state sum~\eqref{e:statesum}. The known
four-dimensional models depend rather strongly on the choice of the
face and edge amplitudes: while some versions of the
Barrett-Crane~$10j$ model have diverging partition sums, others are
dominated by configurations in which almost all spins are zero,
i.e.~configurations which correspond to zero-area
faces~\cite{Baez:2002aw}. Once more, it is important to remember that
even in `old' Regge models in two dimensions, where a comparison with
exact computations in the continuum is
possible~\cite{Knizhnik:1988ak,David:1988hj,Distler:1988jt}, the
continuum limit does not seem to agree with these exact
results~\mbox{\cite{Bock:1994mq,Holm:1994un,Ambjorn:1997ub,Rolf:1998ja}}
(the expectation values of edge lengths do not scale as a power of the
volume when a diffeomorphism invariant measure is used, in contrast to
the exact results).  Therefore, it is far from clear
that~\eqref{e:Zrefined} will lead to a proper continuum limit.

A third proposal is to take a fixed spin foam and to simply define the
model as the sum over all
spins~\cite{Barrett:2000xs,Pfeiffer:2003tx,Freidel:2005qe}; this
proposal differs considerably from both the Regge and dynamical
triangulation approaches.  Considering a fixed foam clearly only makes
sense provided the partition sum is actually independent of the
triangulation of the manifold (or more correctly, one would require
that physical correlators are independent of the triangulation). Such
a situation arises in the three-dimensional Ponzano-Regge model, but
three-dimensional gravity does not contain any local degrees of
freedom. For higher dimensions, the only triangulation independent
models known so far are topological theories, i.e.~theories for which
the local degrees of freedom of the metric do not matter. If one
insists on triangulation independence also for gravity, then one is
forced to add new degrees of freedom to the spin foam models
(presumably living on the edges). In this picture, a change from a
fine triangulation to a coarse one is then compensated by more
information stored at the edges of the coarse triangulation.  This
then also requires (presumably complicated) rules which say how these
new degrees of freedom behave under a move from one triangulation to
another. Note that even when the partition sum is independent of the
refinement of the triangulation, one would probably still want to deal
with complicated cross-sections of foams to describe ``in'' and
``out'' coherent states. At present, there is little evidence that
triangulation independence can be realised in non-topological
theories, or that the problems related to the continuum limit will not
reappear in a different guise.

\section{Predictive (finite) quantum gravity?}

Let us now return to the question as to what can be said about
finiteness properties of spin foam models, and how they relate to 
finiteness properties (or rather, lack thereof!) of the standard perturbative
approach -- after all, one of the main claims of this approach is
that it altogether avoids the difficulties of the standard approach.
So far, investigations of finiteness have focused on the partition sum 
itself. Namely, it has been shown that for a variety of spin foam 
models, the partition sum for a \emph{fixed} spin foam is finite,
\begin{equation}
\label{e:Zfinite}
\sum_{\text{spins $\{j\}$}} Z_\phi\big(\{j\}\big) = \text{finite}\,.
\end{equation}
Even though a given spin foam consists of a finite number of links,
faces, \ldots , divergences could arise in principle because the range
of each spin~$j$ is infinite. One way to circumvent infinite sums is
to replace the group SU(2) by the quantum group $\text{SU}(2)_q$
(which has a finite number of irreps), or equivalently, by introducing
an infinite positive cosmological constant~\cite{Ooguri:1991ni}; in
all these cases the state sum becomes finite.\footnote{The division by
the infinite factor which is required to make the Ponzano-Regge state
sum finite can be understood as dividing out the volume of the group
of residual invariances of Regge models~\cite{Freidel:2002dw}. These
invariances correspond to changes of the triangulation which leave the
curvature fixed. However, dividing out by the volume of this group
does not eliminate the formation of `spikes' in Regge gravity.} A
similar logic holds true in four dimensions and for Lorentzian models,
although in the latter case the analysis becomes more complicated due
to the non-compactness of the Lorentz group, and the fact that the
unitary representations are all infinite
dimensional~\cite{Crane:2001qk}.  Perhaps unsurprisingly, there exist
choices for edge and surface amplitudes in four dimensions which look
perfectly reasonable from the point of view of covariance, but which
are nevertheless not finite~\cite{Baez:2002aw}.

It should, however, be emphasised that the finiteness
of~\eqref{e:Zfinite} is a statement about \emph{\mbox{infrared}}
finiteness. Roughly speaking, this is because the spin~$j$ corresponds
to the `length' of the link, whence the limit of large~$j$ should be
associated with the \emph{infinite volume limit}. In statistical
mechanics, partition functions generically diverge in this limit, but
in such a way that physical correlators possess a well-defined limit
(as quotients of two quantities which diverge). From this point of
view, the finiteness properties established so far say nothing about
the UV properties of quantum gravity, which should instead follow from
some kind of refinement limit, or from an averaging procedure where
one sums over all foams, as discussed above. The question of convergence 
or non-convergence of such limits has so far not received a great
deal of attention in the literature.

This then, in a sense, brings us back to square one, namely the true 
problem of quantum gravity, which lies in the ambiguities associated
with an infinite number of non-renormalisable UV divergences. 
As is well known this problem was originally revealed in a perturbative 
expansion of Einstein gravity around a fixed background, which requires
an infinite series of counterterms, starting with the famous two-loop
result~\cite{Goroff:1985sz,Goroff:1985th,vandeVen:1991gw}
\begin{equation}
\label{e:2loopdiv}
\Gamma^{(2)}_{\text{div}} = \frac{1}{\epsilon}\frac{209}{2880}
\frac{1}{(16\pi^2)^2} \int\!{\rm d}^4 x\, \sqrt{g}\,
C_{\mu\nu\rho\sigma} C^{\rho\sigma\lambda\tau}
C_{\lambda\tau}{}^{\mu\nu}\,.
\end{equation}
The need to fix an infinite number of couplings in order to make the 
theory predictive renders perturbatively quantised Einstein gravity 
useless as a physical theory. What we would like to emphasise here
is that \emph{any} approach to quantum gravity must confront 
this problem, and that the need to fix infinitely many couplings
in the perturbative approach, and the appearance of infinitely many
ambiguities in non-perturbative approaches are really just different
sides of the same coin. 

At least in its present incarnation, the canonical formulation of LQG
does not encounter any UV divergences, but the problem reappears
through the lack of uniqueness of the canonical Hamiltonian. For spin
foams (or, more generally, discrete quantum gravity) the problem is no
less virulent.  The known finiteness proofs all deal with the
behaviour of a single foam, but, as we argued, these proofs concern
the infrared rather than the ultraviolet. Just like canonical LQG,
spin foams thus show no signs of ultraviolet divergences so far, but,
as we saw, there is an \emph{embarras de richesse} of physically
distinct models, again reflecting the non-uniqueness that manifests
itself in the infinite number of couplings associated with the
perturbative counterterms. Indeed, fixing the ambiguities of the
non-perturbative models by \emph{ad hoc}, albeit well motivated,
assumptions is not much different from defining the perturbatively
quantised theory by fixing infinitely many coupling constants `by
hand' (and thereby remove all divergences). Furthermore, even if they 
do not `see' any UV divergences, non-perturbative approaches cannot be 
relieved of the duty to explain \emph{in all detail} how the 2-loop 
divergence (\ref{e:2loopdiv}) and its higher loop analogues `disappear', 
be it through cancellations or some other mechanism. 

Finally, let us remark that in lattice gauge theories, the classical 
limit and the UV limit can be considered and treated as separate issues.
As for quantum gravity, this also appears to be the prevailing view in 
the LQG community. However, the continuing failure to construct viable 
\emph{physical} semi-classical states, solving the constraints even 
in only an approximate fashion, seems to suggest (at least to us) 
that in gravity the two problems cannot be solved separately, but are 
inextricably linked  --- also in view of the fact that the question 
as to the precise fate of the two-loop divergence (\ref{e:2loopdiv}) 
can then no longer be avoided.

\section*{Acknowledgements}

The first part of this paper is based on~\cite{kas_lqg}, written in
collaboration with Marija Zamaklar. We thank Jan Ambj{\o}rn, Herbert
Hamber, Claus Kie\-fer, Kirill Krasnov, Hendryk Pfeiffer, Martin
Reuter and Marija Zamaklar for discussions and correspondence. We are
also grateful to Laurent Freidel for a transatlantic debate that
helped clarify some points in the original version of this review.

\begin{small}
\setlength{\bibsep}{2.5pt}

\begingroup\raggedright\endgroup

\end{small}

\end{document}